\documentclass[amsmath,amssymb,pra,aps,showpacs,superscriptaddress,twocolumn]{revtex4-2}
\usepackage{amsmath,amsfonts,amssymb,amsthm,graphics,graphicx,epsfig,bbm}
\usepackage[colorlinks=true,citecolor=blue,linkcolor=blue,urlcolor=blue]{hyperref}
\usepackage[usenames]{color}
\usepackage{graphicx}
\usepackage{subfigure}
\usepackage{amsmath}
\usepackage{epsfig}
\usepackage{dcolumn}
\usepackage{bm}
\usepackage{color}
\usepackage{times}
\usepackage{epstopdf}
\usepackage{amssymb}
\usepackage{amstext}
\usepackage{latexsym}
\usepackage{hyperref}
\usepackage{amsfonts}
\usepackage{psfrag}
\usepackage{soul,xcolor}
\usepackage[normalem]{ulem}
\usepackage{dsfont}
\usepackage{txfonts}
\usepackage{float}

\newcommand{\ket}[1]{\vert #1 \rangle}
\newcommand{\bra}[1]{\langle #1 \vert}

\newcommand{\Tr}{\mathrm{Tr}}

\begin{document}

\setstcolor{red}

\title{Tunable Non-Markovianity for Bosonic Quantum Memristors}

\author{Jia-Liang Tang }
\affiliation{nternational Center of Quantum Artificial Intelligence for Science and Technology (QuArtist)  and Physics Department, Shanghai University, 200444 Shanghai, China}

\author{Gabriel Alvarado Barrios}
\affiliation{Kipu Quantum, Greifswalderstrasse 226, 10405 Berlin, Germany}

\author{Enrique Solano}
\affiliation{Kipu Quantum, Greifswalderstrasse 226, 10405 Berlin, Germany}

\author{Francisco Albarr\'an-Arriagada}
\email[F. Albarr\'an-Arriagada]{\qquad francisco.albarran@usach.cl}
\affiliation{Departamento de F\'isica, Universidad de Santiago de Chile (USACH), Avenida V\'ictor Jara 3493, 9170124, Santiago, Chile.}
\affiliation{Center for the Development of Nanoscience and Nanotechnology 9170124, Estaci\'on Central, Santiago, Chile.}

\begin{abstract}
We study the tunable control of the non-Markovianity of a bosonic mode due to its coupling to a set of auxiliary qubits, both embedded in a thermal reservoir. Specifically, we consider a cavity mode coupled to auxiliary qubits described by the Tavis-Cummings model.  As a figure of merit, we define the dynamical non-Markovianity as the tendency of a system to return to its initial state, instead of evolving monotonically to its steady state. We study how this dynamical non-Markovianity can be manipulated in terms of the qubit frequency. We find that the control of the auxiliary systems affects the cavity dynamics as an effective time-dependent decay rate. Finally, we show how this tunable time-dependent decay rate can be tuned to engineer bosonic quantum memristors, involving memory effects that are fundamental for developing neuromorphic quantum technologies.
 \end{abstract}

\maketitle


\section{Introduction}
In open quantum systems, the Markovian approximation is widely used due to its mathematical simplicity and the good description of the phenomenology observed in the lab. The Markovian approximation, from a pedagogical perspective, considers that the state of the reservoir is not correlated at different times, which can be interpreted as a memoryless bath~\cite{ Li2019EPL, Li2020EPL}. Nevertheless, in many real-world systems, memory effects emerge, causing non-trivial dynamics, where transport effects in biological systems are one of the most paradigmatic ones~\cite{Lee2007Science, Chin2010NewJPhys, Fleming2011NewJPhys}. It suggests that the manipulation of non-Markovian systems and the control of the open system dynamics is important for several applications, such as quantum metrology~\cite{Alex2012PhysRevLett, Nicolas2012PhysRevA}, quantum simulation~\cite{Sweke2016PhysRevA, Julio2011Nature}, and quantum memdevices~\cite{Pfeiffer2016SciRep, Michele2022NatPhoton, Sanz2018APLPhotonics, Shevchenko2016PhysRevAppl, Norambuena2022PhysRevAppl, Kumar2021PhysRevA}. In this context, the manipulation of the non-Markovianity looks promising for implementing new technological devices, particularly for quantum memristive systems useful for the realization of neuromorphic computing at a quantum level~\cite{Pershin2012IEEE, Pehle2022PhysRevE, Xu2021PhysRevB}.  

On the other hand, the definition and quantification of the non-Markovianity in quantum systems is still an open question, as can be checked in the recent literature~\cite{Vega2017RevModPhys, Rivas2014RepProgPhys, Breuer2012JPhysB, Breuer2009PhysRevLett, Rivas2010PhysRevLett}. Nevertheless, there are two widely accepted cases by the scientific community. The first is based on the distinguishability of a quantum system~\cite{Breuer2009PhysRevLett} under a dissipative evolution. This definition considers that if a system interacts with a Markovian environment, the system's information will flow unidirectionally to the environment. It means that the system loses distinguishability between different initial states during the dynamics. In other words, the system monotonically forgets the initial condition. Oppositely, in a system with a non-Markovian environment, the distinguishability between two evolutions with different initial conditions will increase at some time, recovering the lost information from the environment. This definition allows us to quantify the non-Markovianity of a dissipative channel by the sum of the regions where some distance measure increases in time for some pair of initial conditions. The second definition of non-Markovianity is based on entanglement with the auxiliary system~\cite{Rivas2010PhysRevLett}. The system and the auxiliary system are in maximum entanglement at the initial moment. If the entanglement gradually decreases as the system evolves, the system is in a Markovian environment. Now, if the entanglement does not decrease monotonically, it means that it increases for some time, then the system interacts with a non-Markovian environment. Even if a method exists to quantify the non-Markovianity of a channel in both cases, the degree of non-Markovianity involves an optimization process over all the possible evolutions (initial state), quantifying the increments of the distinguishability or entanglement as a cost function.

Recently, the non-Markovian dynamics has been actived researched both in theory and experiment, driven by the wide interest in quantum technologies~\cite{Luchnikov2020PhysRevLett, Victor2018NewJPhys, Liu2011NaturePhys, Bernardes2015SciRep, Li2022PhysRevLett, Garcia2020npjQuantumInf, Chen2022npjQuantumInf}. For example, from a theoretical point of view, with a harmonic oscillator coupled to both non-Markovian and Markovian baths, many characteristics of the system can be explored, like spectral properties~\cite{Victor2018NewJPhys}.  In experiments using an all-optical experiment, the transitions between Markovian and non-Markovian regimes can be reached, controlling the information backflow of the system~\cite{Liu2011NaturePhys} as well as the observation of the called weak non-markovinity regime~\cite{Bernardes2015SciRep}. 

In this article, we focus on the dynamical non-Markovianity (DnM), which means the degree of non-Markovianity presented in a given dynamics. Specifically, we will focus on a system composed of a cavity mode (main system) coupled to a set of qubits (auxiliary systems) described by the Tavis-Cummings model~\cite{Retzker2007PhysRevA, Jager2022PhysRevLett} embedded in a Markovian bath. We are interested in studying the DnM that arises in the main system dynamics by tracing the auxiliary qubits, creating a tunable bosonic quantum memristor. Also, we will explore how the DnM can be manipulated by external control over the auxiliary systems. We find that by tuning the energy gap of the set of qubits, we can simulate a time-dependent decay rate in the cavity going from a regime with maximal DnM and another with minimal DnM and Markovian evolution. This tunable dynamical Non-Markovianity, allow us to define variables that follow a memristive behavior, obtaining an experimental feasible, scalable and general framework to implement switchable memory devices useful for neuromorphic quantum computing.


\begin{figure}[t]
\includegraphics[width=0.8\linewidth]{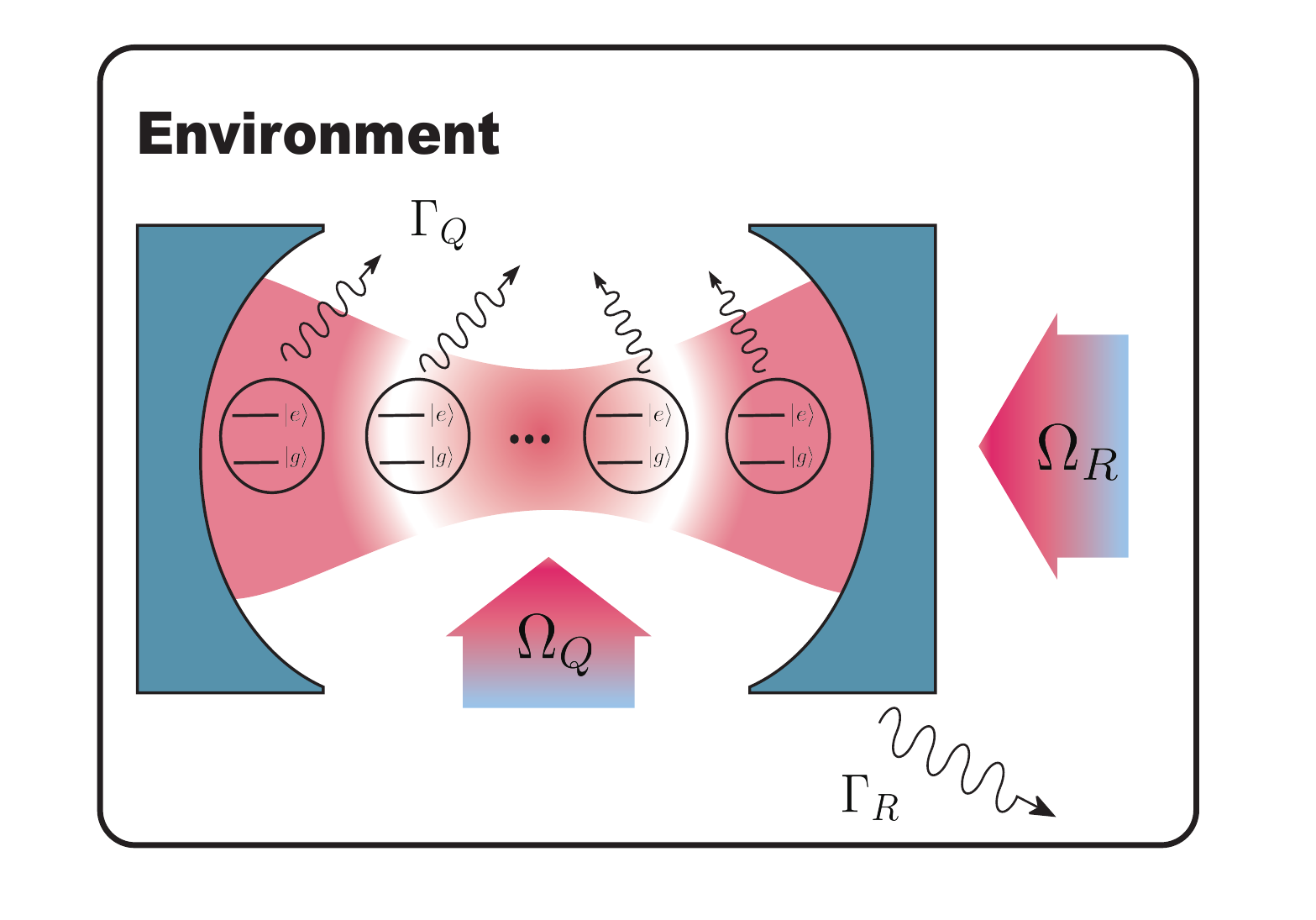}
\centering
\caption{Diagram of the model: a cavity (bosonic mode) coupled to a set of qubits embedded in a Markovian reservoir. Each auxiliary qubit can be dynamically tuned and the cavity can be classically driven.}
\label{Fig01}
\end{figure}   

\section{\label {model}Model and Methods}
We consider a system consisting of a single bosonic mode (resonator) coupled to a set of $n$ qubits in contact with a thermal reservoir at zero temperature as shown in Fig.~\ref{Fig01}. The interaction between the qubits and the resonator is described by the Tavis-Cumming model
\begin{equation}
\hat{H}_{TC}=\hat{H}_R+\hat{H}_Q+\hat{H}_{R-Q}
\label{Hamiltonian}
\end{equation}
where
\begin{eqnarray}
\hat{H}_R&=&\hbar\omega_R\hat{a}^{\dagger}\hat{a},\nonumber\\
\hat{H}_Q&=&\frac{\hbar}{2}\sum_{j=1}^{n}\omega_{Q}\hat{\sigma}_{z,j},\nonumber\\
\hat{H}_{R-Q}&=&\hbar\sum_{j=1}^{n}g(\hat{\sigma}_{j}^{-}\hat{a}^{\dagger}+\hat{\sigma}_{j}^{+}\hat{a})),
\end{eqnarray}
are the Hamiltonians for the bosonic mode, the qubits, and the resonator-qubits interaction, respectively. Here, $\omega_R$, $\omega_{Q}$, $g$, and $\hbar$ represent the resonator frequency, the qubit frequency, the qubit-resonator coupling strength, and the Planck constant. The operator $\hat{a}(\hat{a}^{\dagger})$  is the annihilation(creation) operator for the bosonic mode, $\hat{\sigma}_{z,j}$ is the Pauli $z$ matrix for the $j$th qubit, and $\hat{\sigma}_j^{-(+)}$ is the lowering(raising) operator for the $j$th qubits. In order to ensure the validity of our model, we consider $\omega_Q/\omega_R\sim 1$ and $g/\omega_R< 0.1$.  From now we will consider $\hbar=1$.

We consider that the total system undergoes Markovian dynamics described by the following master equation,
\begin{eqnarray}
&&\dot{\rho}(t)=-i[\hat{H}_{TC},\rho(t)] + \sum_{j=0}^n\Gamma_j\mathcal{D}[\mathcal{\hat{O}}_{j}]\rho,
\label{Eq03}
\end{eqnarray}
with
\begin{equation}
\mathcal{D}[\mathcal{\hat{O}}_{j}]\rho=\mathcal{\hat{O}}_{j}\rho\mathcal{\hat{O}}_{j}^{\dagger}-\frac{1}{2}\{\mathcal{\hat{O}}_{j}^{\dagger}\mathcal{\hat{O}}_{j},\rho\},
\end{equation}
where $\mathcal{\hat{O}}_{0}=a$ and $\mathcal{\hat{O}}_{j}=\sigma_j^-$ for $j>0$ and $\Gamma_j$ is the decay rate of the $j$th channel. We are interested only in the dynamics of the resonator, thus we focus on its reduced state by tracing out the qubits, $\rho_R(t)=\Tr_Q(\rho(t))$. In this way, the set of qubits act as an auxiliary system that introduces non-Markovian properties to the dissipative evolution of the resonator. We want to characterize the degree of non-Markovianity of a particular evolution of our system (the resonator) determined by its initial state. We look for a figure of merit that can be understood as a degree of non-Markovianity of the particular dynamics of the system that result from a given initial condition. To this end, we notice that when the dynamics of the system are Markovian and purely dissipative, then its quantum state will monotonically approach the corresponding steady state of the environment. We can characterize this behavior by calculating the trace distance between the instantaneous state of the system and the steady state of the evolution,
\begin{equation}
D_S(\rho_R(t))=\frac{1}{2}|\rho_R(t)-\rho_{SS}| ,
\label{trace_distance}
\end{equation}
where the subindex $S$ denotes that the trace distance is taken with respect to the steady state. For a Markovian evolution, this quantity will decrease monotonically to zero \cite{Breuer2009PhysRevLett}, where ${|\rho| = \Tr[\sqrt{\rho^\dagger \rho}]}$ and $\rho_{SS}$ is the steady state of the system. In our case, the temperature of the environment is zero and therefore $\rho_{SS}=\ket{0}\bra{0}$. Now, the quantity $D(t)$ allows us to detect when the evolution deviates from Markovian behavior whenever it is no longer monotonically decreasing. Therefore, we can characterize the non-Markovianity of a particular system evolution by considering all the time intervals with non-monotonic behavior. In this way, we define the DnM as
\begin{equation}
\mathcal{N}_D=\int_{\zeta>0}\zeta(t) dt ,
\end{equation}
where $\zeta(t)=(d/dt)D_S(\rho_R(t))$, for an evolution long enough to reach the steady state. We note that this definition is closely related to the non-Markovianity measure for dissipative channels based on distinguishability~\cite{Breuer2009PhysRevLett}. However, our definition considers only the dynamics under study and not an optimization over all the initial conditions. In our system, the qubit-resonator coupling is the factor that introduces non-Markovian behavior into the resonator dynamics due to information backflow from the set of qubits. In the next section, we will characterize how a given configuration of the set of qubits affects the behavior of the DnM for such bosonic quantum memristor.


\begin{figure}[b!]
\centering
\includegraphics[width=\linewidth]{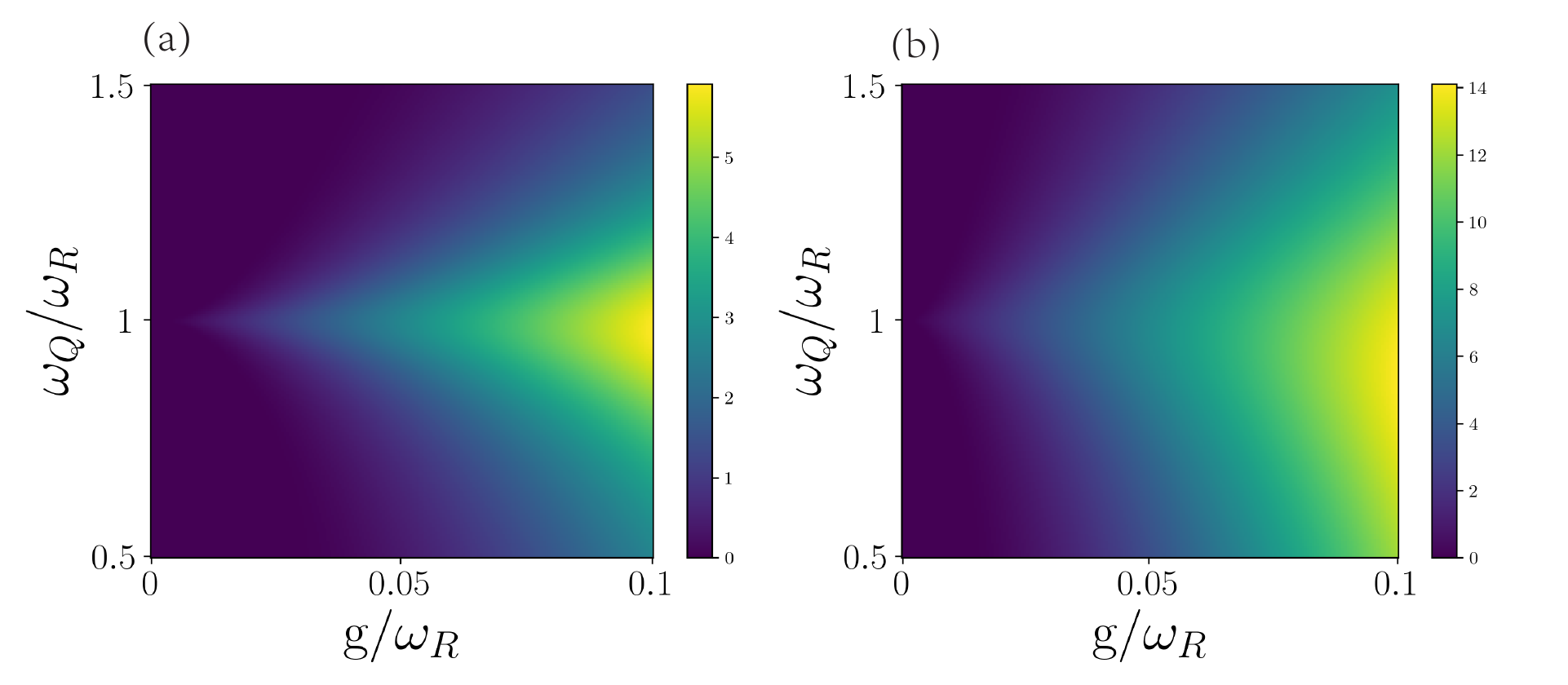}
\caption{Dynamical non-Markovainity of the resonator. (a) one-qubit case. (b) five-qubit case.  In both cases, the decay rate of qubit and resonator is $\Gamma_Q=\Gamma_R=0.005$. We consider all qubits are in resonance ($\omega_Q=\omega_R$), the coupling strength $g/\omega_R\in [0, 0.1]$ and the initial state $|\psi_0\rangle=|1_R0_Q\rangle$.}
\label{Fig02}
\end{figure}

\begin{figure}[t]
\centering
\includegraphics[width=0.85\linewidth]{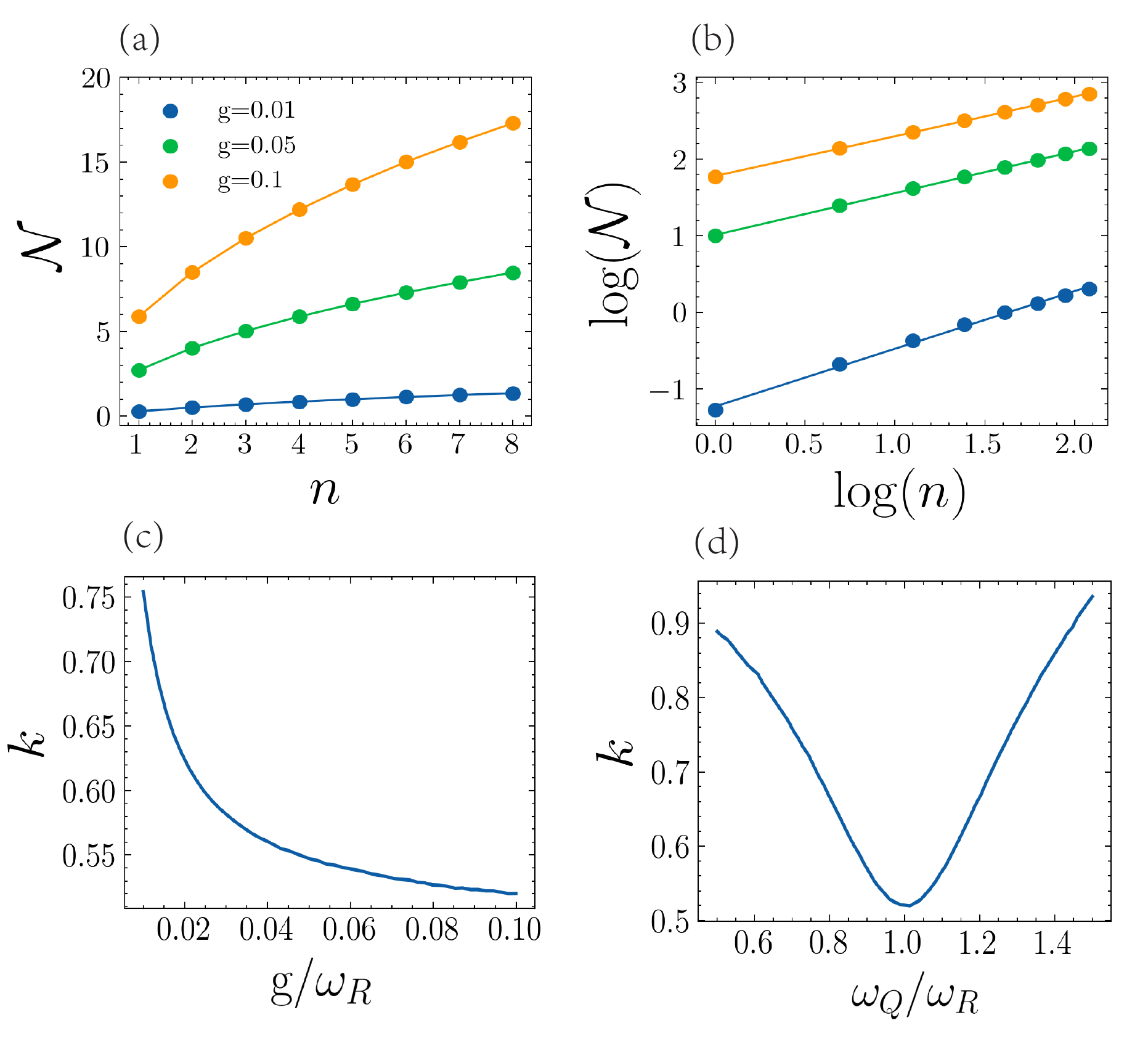}
\caption{(a) The DnM of the resonator in terms of the number of qubits $n$. (b) the log-log plot of DnM and the number of qubits. For (a) and (b), we consider three cases $g=0.01, 0.05, 0.1$, with the resonant condition $\omega_Q=\omega_R=1$. (c) The exponent $k$ of the power-law dependence as a function of the coupling strength $g/\omega_R$, with qubit and resonator in resonance. (d) The exponent $k$ of the power-law dependence as a function of the frequency of the qubit $\omega_{Q}/\omega_R$ and a fixed coupling strength $g/\omega_R=0.05$. For all cases, we consider decay rates $\Gamma_Q=\Gamma_R=0.005\omega_R$, the initial state of the resonator $|\psi_0\rangle=|1_R\rangle$, and all the qubits initialized in the ground state. }
\label{Fig03}
\end{figure}

\section{\label{result}Results}
\subsection{Dynamical non-Markovianity (DnM)}

For our first case, we will focus on the resonator interacting with one qubit (Jaynes-Cummings model) and interacting with $n=5$ qubits. We will analyze how the DnM depends on the qubit frequency and coupling strength.  It is important to mention that the set of auxiliary qubits is always initialized in the ground state in order not to introduce energy into the resonator since it would undermine the interpretation of the DnM. First, we consider the initial state $|\psi_0\rangle=|1_R0_Q\rangle$. In Fig. \ref{Fig02}~(a), we show the DnM of the resonator when varying the coupling strength $g/\omega_{\textrm{R}}$ and the frequency ratio $\omega_{\textrm{Q}}/\omega_{\textrm{R}}$. We can see that the DnM is largest when qubit and resonator are in resonance and when $g$ increases. Notice that for larger values of $g/\omega_{\textrm{R}}$, the qubit-resonator detuning can yield significant values of DnM. Figure \ref{Fig02}~(b) shows the DnM for the case of five auxiliary qubits. We can observe that the effect of enlarging the set of auxiliary qubits is relaxing the resonance condition and increasing the value for the DnM.

This behavior is to be expected since the resonance condition allows for maximal information transfer and information backflow due to the complete Rabi oscillations (in the case of $n=1$). In addition, the coupling strength $g/\omega_{\textrm{R}}$ is related to the speed of the information transfer and information backflow. Then, for small $g$ (slow information transfer), a stronger resonance condition is needed to have information backflow before the system reaches the stationary state. If $g$ increases the communication between the auxiliary qubits and the bosonic mode is faster, and a more relaxed resonance condition will still have information backflow. Increasing the number of qubits, increases the channels for information backflow, which leads to larger values of DnM even at higher detuning.

Next, we study the scaling of the DnM with the number of qubits under fixed conditions. In Figure~\ref{Fig03}, we study $\mathcal{N}_{D}$  as a function of the number of qubits (until $n=8$) for different coupling strengths when the resonator is initialized with one excitation. 
Figure~\ref{Fig03}~(a) shows how $\mathcal{N}_{D}$  scales with the number of particles ($n$), with a monotonically increasing behavior reminiscent of a power law. In Fig.~\ref{Fig03}~(b),  we do a log-log plot of the quantities of Figure~\ref{Fig03}~(a) which confirms the power-law dependence. In all instances the $R^2$ coefficient is larger than $0.995$ which means that the scaling of the DnM is very well approximated by
\begin{equation}
\mathcal{N}_D\propto n^k,
\end{equation}
where the exponent $k$ depends on the coupling strength $g/\omega_R$ and the qubit frequency $\omega_Q/\omega_R$, as shown in Fig.~\ref{Fig03}~(c) and Fig.~\ref{Fig03}~(d), respectively. We can see that when the DnM is maximal, that is, for large $g$ and $\omega_Q=\omega_R$, the value of $k$ is minimum. This is a finite size effect of the auxiliary system as we would expect that if the number of qubits increases to the thermodynamic limit they would induce Markovian dynamics for the resonator.

 \begin{figure}[t]
\centering
\includegraphics[width=\linewidth]{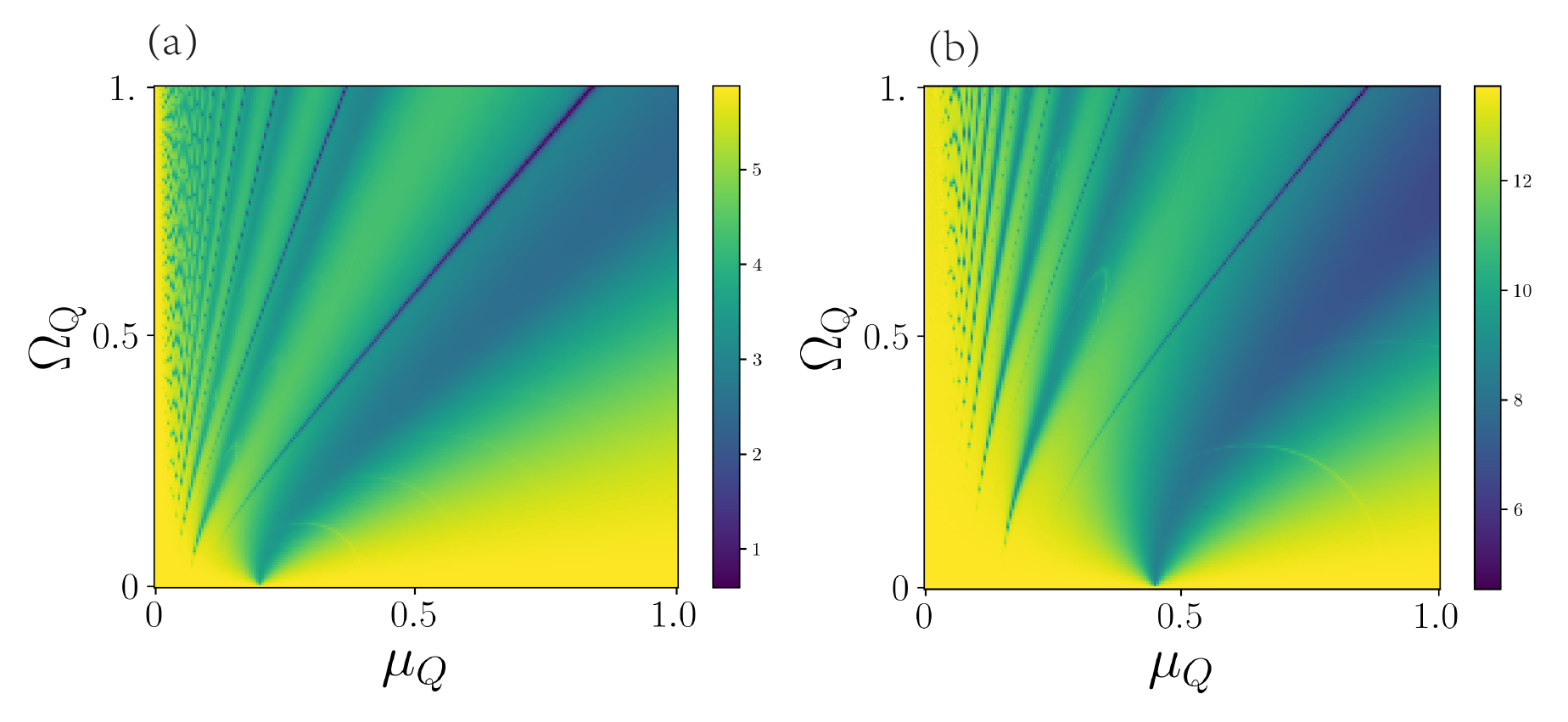}
\caption{The non-Markovainity of the resonator. (a) one-qubit case. (b) five-qubit case.  Parameters: In both cases, the decaying rate of qubit and resonator is $\Gamma_Q=\Gamma_R=0.005\omega_R$.  The driving frequency of qubit $\mu_Q/\omega_R\in[0,1]$, the driving amplitude of the qubit $\Omega_Q/\omega_R\in[0,1]$, the qubit frequency $\omega_Q/\omega_R=1$, the coupling strength $g/\omega_R=0.1$. The initial state is  $|\psi_0\rangle=|1_R0_Q\rangle$.}
\label{Fig04}
\end{figure}

We have seen that the DnM of the resonator strongly depends on the parameters of the set of auxiliary qubits. It is then interesting to consider whether we can have dynamic control over the DnM by manipulating the set of auxiliary qubits. In what follows, we apply a driving term in the $z$-direction to the set of auxiliary qubits in order to dynamically modulate the qubit gap and control the degree of DnM in the evolution. The driving is chosen so that it does not introduce energy into the qubits which could excite the resonator and be interpreted as information backflow by the DnM. This situation is described by the following Hamiltonian
 \begin{equation}
\hat{H} =H_{TC}+\hbar\Omega_Q\sum_j^n\sin(\mu_Q t)\hat{\sigma}^z_j,
\label{Eq08}
\end{equation}
where $H_{TC}$ is the Hamiltonian of eq.(\ref{Hamiltonian}),  $\Omega_Q$ and $\mu_Q$ are the amplitude and frequency of the driving over the qubits, respectively. Notice that we consider that each qubit is driven by the same signal. 

 \begin{figure}[t]
\centering
\includegraphics[width=\linewidth]{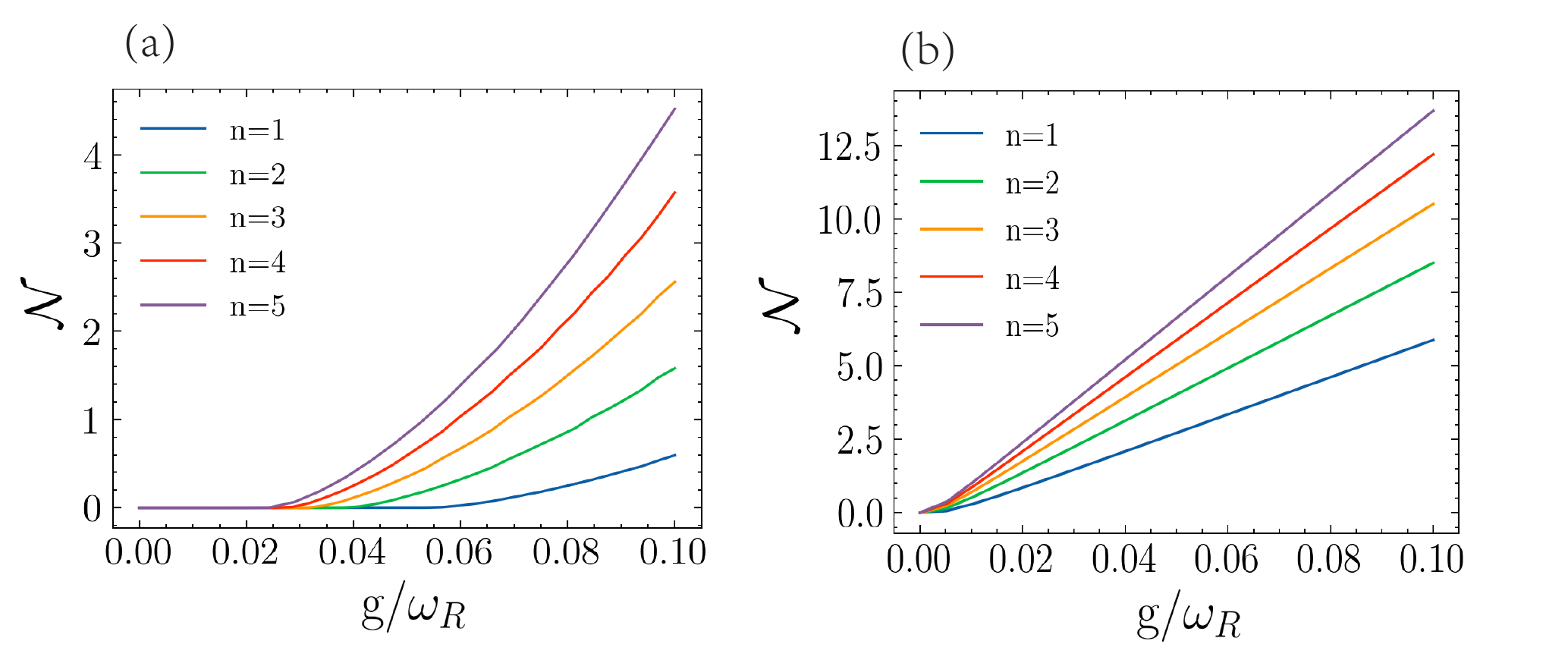}
\caption{The minimum (a) and maximum (b) of DnM for the resonator with a different number of auxiliary qubits. In both cases, the decaying rate is  $\Gamma_Q=\Gamma_R=0.005\omega_R$.  The frequency of qubit $\omega_Q/\omega_R=1$, and the initial state is of resonator $|\psi_0\rangle=|1_R\rangle$. The qubits are all initialized in the ground state.}
\label{Fig05}
\end{figure}
We numerically calculate $\mathcal{N}_{D}$ for different values of the driving frequency and amplitude which we show in Fig. \ref{Fig04}. In Fig. \ref{Fig04} (a) we show the case of one auxiliary qubit. Here, there is non-zero DnM over the whole range of parameters, however it is interesting to notice the dark lines that are spanned from near the origin where the DnM is almost completely suppressed. A similar behavior occurs when we increase the number of qubits, as is shown in Fig. \ref{Fig04} (b) for the five qubits case, where the DnM is suppressed over thin lines in the frequency/amplitude plane. Although the suppression is not as strong as in the one-qubit case, these lines show significant decrease in the DnM. This indicates that by modulating either the frequency or the amplitude of the driving we can enhance or suppress the DnM of the resonator.  

Similarly, we study the DnM in terms of the coupling strength for different number of qubits in the auxiliary system. Here, we will consider two cases, the parameters of the driving that yield the maximum and minimum DnM. In Fig. \ref{Fig05} (a), we plot the minimal DnM for different coupling strength $g$. We can see that up to $g=0.02$ we can essentially completely suppress the non-Markovian behavior by a suitable choice of driving parameters. Increasing the number of qubits decreases the necessary value of coupling strength that allows for completely supressed DnM. On the other hand, in Fig. \ref{Fig04} (b), we plot the maximal DnM for different coupling strength $g$. Here we can see that $\mathcal{N}_{D}$ has linear dependence on the coupling strength except for a small range around zero. From these results we have that provided we choose a suitable value of the coupling strength, we can switch between Markovian and non-Markovian dynamics for the resonator by just controlling the auxiliary set of qubits. 

\begin{figure}[t!]
\centering
\includegraphics[width=0.75\linewidth]{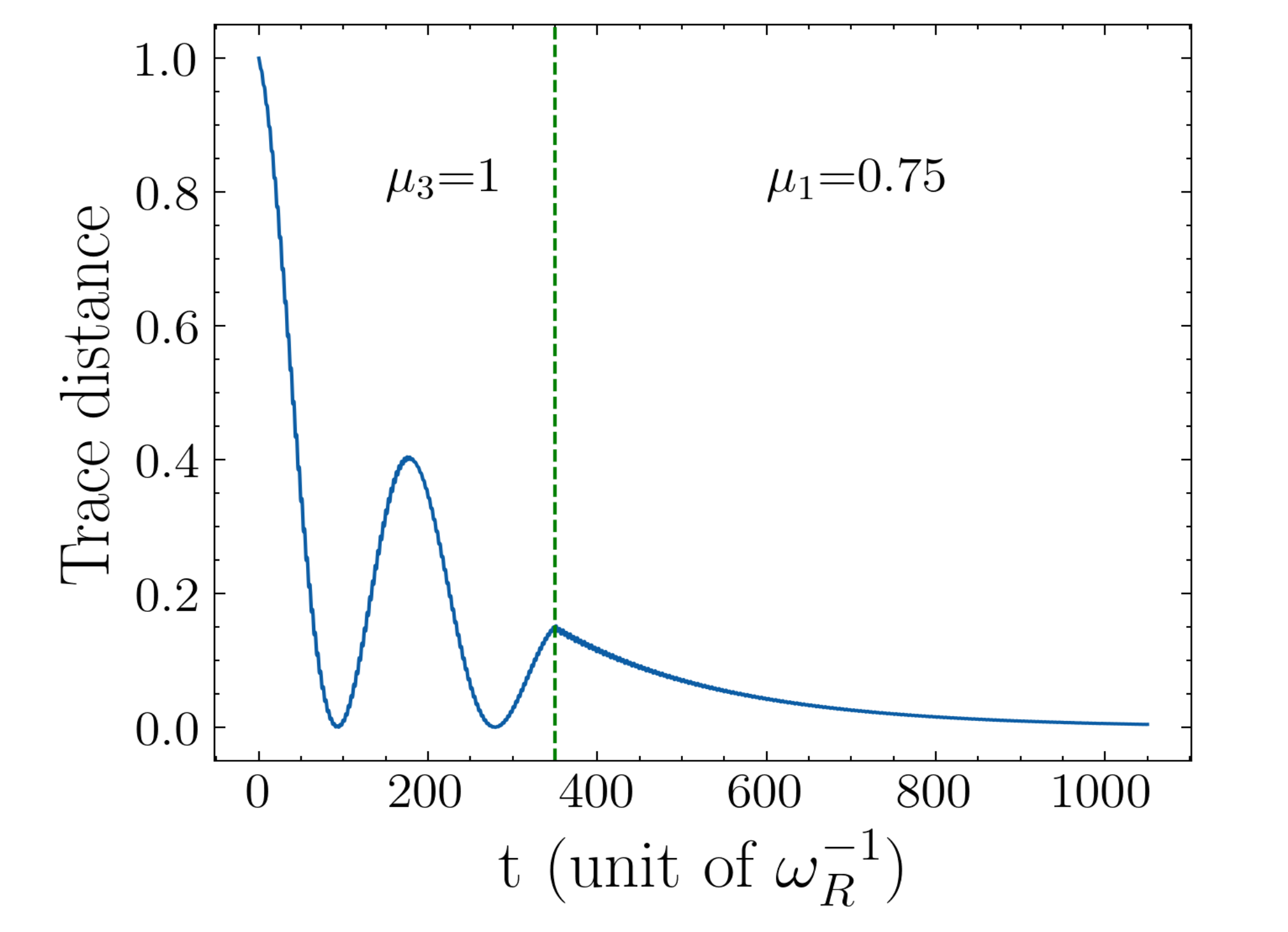}
\caption{Transition from non-Markovian to Markovian dynamics by changing the driving frequency over the auxiliary qubits. Parameters: the driving amplitude of qubit is $\Omega_q=0.5$,  the number qubit $n=1$, decaying rate is $\Gamma_Q=\Gamma_R=0.005$, frequency $\omega_Q=\omega_R$, coupling strength $g=0.05$.}
\label{Fig07}
\end{figure}

In Fig.~\ref{Fig07} we show how we can dynamically switch the non-Markovian behavior on and off by just changing the driving frequency of the qubits. Here, we plot the trace distance as a function of time. At the start of the evolution, we choose a driving frequency that yields maximum non-Markovianity ($\mu_Q=\Omega_R$), later at $t=350\omega_R^{-1}$ we switch the driving frequency to $\mu_{Q} = 0.75\omega_R$ which yields minimum DnM. As can be seen in the figure, at $t=350\omega_R^{-1}$, the trace distance switches from non-monotonic to monotonically decreasing behavior, which characterizes Markovian evolution. 

\subsection{Time-dependent decay rate}
The observed memory effect can be understood as the system effectively interacting with an environment with a time-dependent decay rate, which becomes negative during some time intervals, favoring the information back-flow~\cite{Hu2015AnnPhys}.  To understand this statement, consider a resonator, with state $\tilde{\rho}$, undergoing dissipative dynamics with a time-dependent decay rate and without any interaction with an auxiliary system, the system is then described by the following master equation
\begin{equation}
\dot{\tilde{\rho}}(t) = -i[H,\tilde{\rho}] + \Gamma(t)\left(a\tilde{\rho} a^\dagger-\frac{1}{2}\{\tilde{\rho},a^\dagger a \}\right)
\label{Eq09}
\end{equation}
where $H = \hbar\omega_R a^\dagger a $, and $\Gamma(t)$ is a time-dependent decay rate that can be negative. Here, $\tilde{\rho}$ represents the state of the resonator undergoing dynamics as described above, and is different from $\rho_{\textrm{R}}$ which is the reduced state of the resonator as described by Hamiltonian~(\ref{Hamiltonian}). For $\Gamma(t)>0$, the energy of the resonator dissipates to the environment, meaning that the information in the resonator is continuously lost. Meanwhile, for  $\Gamma(t)<0$ there is energy entering the resonator, giving place to information back-flow and therefore to a non-Markovian process.
\begin{figure}[t]
\centering
\includegraphics[width=\linewidth]{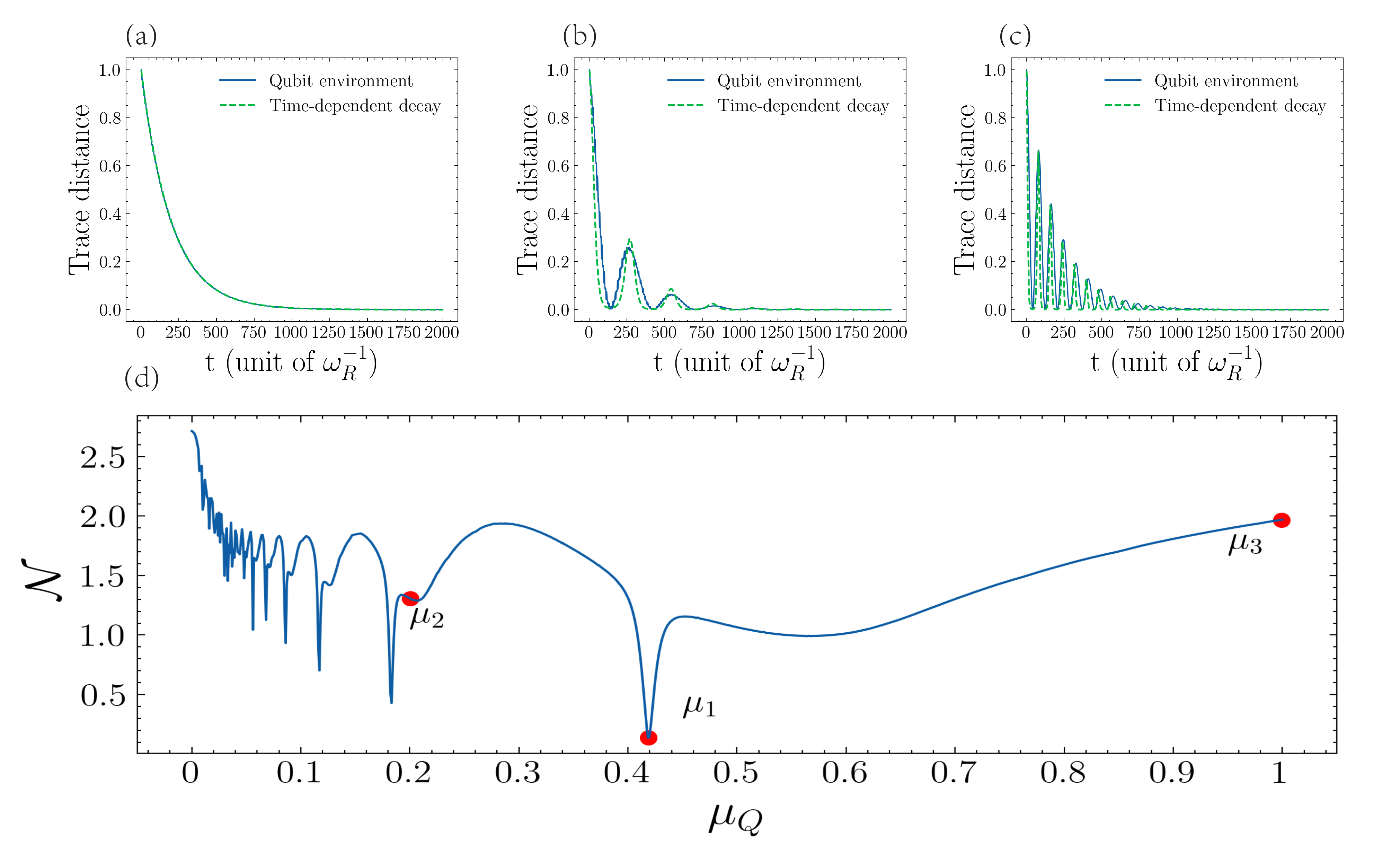}
\caption{Trace distance of the resonator and the DnM under different driving frequencies. Top, (a)  the blue line, the frequency of driving $\mu_1=0.419$, the green line, the resonator's decaying rate is constant $\Gamma(t)=0.005$. (b) the blue line,  the frequency of driving $\mu_2=0.20$, the green line, the resonator's decaying rate is $\Gamma_1(t)=0.05\left(\sin(0.023t)+0.09\right)$. (c) the blue line, the frequency of driving $\mu_3=1$. the green line, the resonator's decaying rate is $\Gamma_1(t)=0.25\left(\sin(0.079t)+0.021\right)$. Bottom, (d) the DnM of resonator in different driving frequency $\mu_Q\in(0, 1)$. Parameters:  the number qubit $n=1$, decaying rate is $\Gamma_Q=\Gamma_R=0.005$, frequency $\omega_Q=\omega_R$, coupling strength $g=0.05$, the driving amplitude $\Omega_Q=0.5\omega_R$.}
\label{Fig06}
\end{figure}

We consider the time-dependent decay rate parametrized as $\Gamma_1(t)=A\left(\sin(Bt)+C\right)$. Notice that the master equation of Eq.~(\ref{Eq09}) has the same steady state as that of our original system in Eq.~(\ref{Eq03}). Therefore, for a given dynamics induced by the set of auxiliary qubits, we can find the closest non-Markovian dynamics corresponding to negative decay rate by finding $A$, $B$, and $C$ that minimize the difference of trace distance $\vert D(\rho_R(t)) - D(\tilde{\rho}(t)) \vert$. In Fig. \ref{Fig06}, we plot $D_S(\rho_R(t))$ and $D_S(\tilde{\rho}_{\textrm{opt}}(t))$ where $\rho_R(t)$ is for one qubit case and $\tilde{\rho}_{\textrm{opt}}(t)$ is the resonator evolved with the optimal parameters for the decay rate.

We consider 3 cases, in Fig. \ref{Fig06} (a), the effective decay rate is time-independent $\Gamma=0.005$ corresponding to Markovian behavior; whereas for time-dependent decay rate we have in Fig. \ref{Fig06} (b) where $\Gamma(t)=0.05[\sin(0.023t)+0.09]$ and Fig. \ref{Fig06} (c) where $\Gamma(t)=0.25[\sin(0.079t)+0.021]$. Finally, Fig. \ref{Fig06} (d) shows the DnM as a function of the qubit-driving frequency where it displays the qubit driving frequency corresponding to Fig.~\ref{Fig06} (a) - (c). We can see that for time-dependent decay the behavior of both trace distance is very similar, which means that the set of auxiliary qubits is inducing highly non-Markovian dynamics.

Finally, it is interesting to study how this simulated time-dependent decay rate can affect the response of the cavity over external driving, in order to control the memristive properties of the dynamics.

\begin{figure}[t]
\centering
\includegraphics[width=0.9\linewidth]{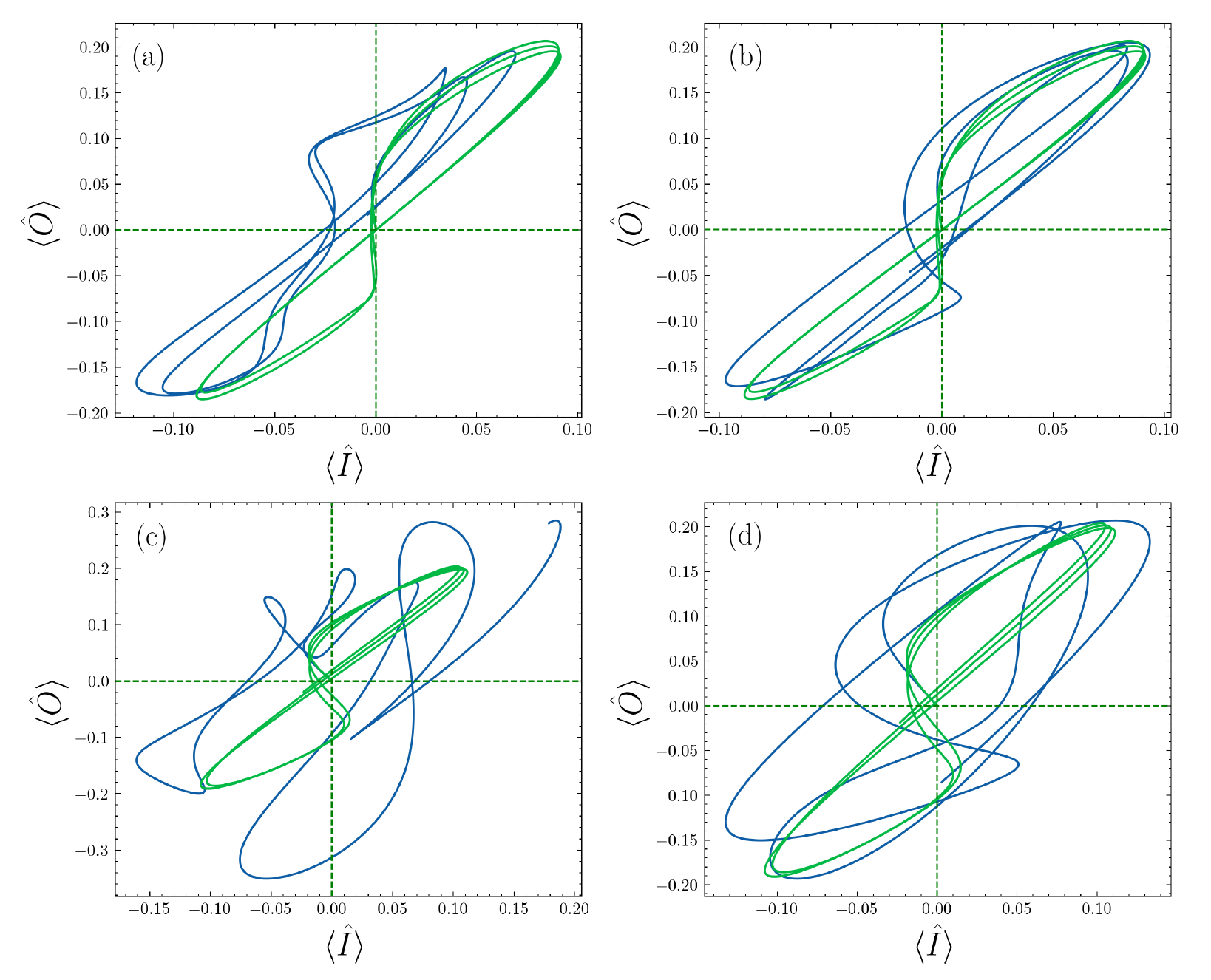}
\caption{Memristive behavior, the green line shows the dynamics when the auxiliary qubits are not driven and off-resonant and the blue curve is when we add a driving over the auxiliary qubits. (a) larger-DnM case, the number of qubits $n=1$, driving frequency $\mu_c=1$. (b) medium-DnM case, the number of qubits $n=1$, driving frequency $\mu_c=0.2$. (c) larger-DnM case, the number of qubits $n=5$, driving frequency $\mu_c=1$. (d) medium-DnM case, the number of qubits $n=5$, driving frequency $\mu_c=0.2$. Parameters: the driving amplitude of qubit is $\Omega_q=0.5$, decaying rate is $\Gamma_Q=\Gamma_R=0.005$, frequency $\omega_Q=\omega_R$, coupling strength $g=0.05$, the driving amplitude of cavity $\Omega_c=0.2$, frequency $\mu_c=0.5$.} 
\label{Fig08}
\end{figure}

\subsection{Bosonic quantum memristor}
One interesting application of our results is to induce memristive behavior into the bosonic mode, which can be tuned by the set of auxiliary qubits. In Ref. \cite{Salmiletho2017SciRep}, it was shown that a kind of time-dependent decay rate produces a quantum memristor, which could be reached in a superconducting circuits platform. Later, in Ref. \cite{Guo2022PhysRevAppl}, a memristive dynamics was obtained in a quantum computer by the simulation of a non-Markovian bath. In this line, we analyze the response of the cavity under an external driving, obtaining a Hamiltonian of the form:
 \begin{equation}
\hat{H} =H_{TC}+\hbar\Omega_Q\sum_j^n\sin(\mu_Q t)\hat{\sigma}^z_j + F(t) (a + a^{\dagger}).
\label{Eq08}
\end{equation}

Now, if we define the variables $I=-\langle i(a - a^{\dagger})\rangle$ and $O=\langle \dot{N} \rangle + \alpha\langle N \rangle $, with $\alpha$ a constant. If we consider $\alpha=\Gamma_c$, it is the natural decay rate of the cavity, we have that 
\begin{equation}
O = F(t)I + \mathcal{G}(t),
\end{equation}
for more details see appendix \ref{appdex1}. The function $\mathcal{G}(t)$ depends on the DnM of the system, which means that we can control the memristive relation 
\begin{equation}
O = F(t)I,
\end{equation}
by controlling the value of $\mathcal{G}(t)$. Now, if we choose $F(t)=\Omega_c[1-\sin(\cos\mu_ct)]$, it is possible to obtain the typical pinched hysteresis loop that characterizes a quantum memristor.

This situation is shown in Fig. \ref{Fig08}, where we obtain the pinched hysteresis loop (green curve), in a similar way that it is obtained for previous proposals of quantum memristos~\cite{Sanz2018APLPhotonics,Norambuena2022PhysRevAppl,Salmiletho2017SciRep} as a signature of memristive behavior. It is interesting to note that such bosonic quantum memristor dynamics appear when the auxiliary qubits are not driven and off-resonant with the cavity, which means that in an effective way, the qubits are decoupled from the cavity. In contrast, we can observe that when we drive the qubits, the memristive behavior can be destroyed for different cases, obtaining a way to go from memristive dynamics to non-memristive dynamics. It means that we also can control the memory properties induced by the decay rate in the cavity, which can be helpful in neuromorphic computing, considering that the proposed system can be implemented in many platforms like trapped ions, optical devices, and superconducting circuits, among others. We also need to remark that our proposal can work as a switchable bosonic quantum memristor. This suggests that our formalism allows implementing devices with controllable and switchable memory properties, only by tuning the energy gap of auxiliary qubits. This proposal opens the door for the experimental implementation of memristive devices, providing a general, platform-free, and scalable model for the next generation of neuromorphic quantum computing technology.

\section{\label{Conclusion}Conclusions}

We consider a cavity coupled to a set of auxiliary qubits, which induce a controllable dynamical non-Markovianity (DnM). We show that by dynamical tuning of the energy gap of the auxiliary qubits, we can go from high DnM to low DnM, which can be considered as an effective time-dependent decay rate. We also show that the induced DnM in the cavity mode follows a power-law dependence with the number of auxiliary qubits. Finally, we show as an application that we can define memristive dynamics in the bosonic mode, which can be switched off by controlling the energy gap of the auxiliary qubits. This means that we can control the memristive dynamics in the cavity by external control over the auxiliary system, obtaining a switchable bosonic quantum memristor. These results provide a general protocol to obtain controllable bosonic quantum memristors which can be useful in neuromorphic quantum computing. This proposal is experimentally feasible since it only uses a bosonic mode coupled to a set of qubits, a ubiquitous setup in hardware platforms like trapped ions, superconducting circuit quantum electrodynamics, and atomic devices.

\onecolumngrid

\appendix
\section{The derivative of expectation the number of photons}

\begin{equation}
\frac{d\langle \hat{a}^\dagger\hat{a} \rangle}{dt}={\Tr\left[-\frac{i}{\hbar}\left[H, \rho \right]\hat{a}^\dagger\hat{a}  \right]}
+\Tr\left[\mathcal{L}(\rho)a^\dagger a \right] = \mathcal{S}_1 + \mathcal{S}_2
\end{equation}
we set $\hbar=1$, $\mathcal{S}_1$ and $\mathcal{S}_2$ is

\begin{eqnarray}
\mathcal{S}_1&=&-i\Tr\left[ \left[ \omega_ca^\dagger a + \omega_q\sum\limits_{j=1}^{n}\sigma_j^z + \sum\limits_{j=1}^{n} g(a^\dagger \sigma_j + a\sigma_j^+)+ \sum\limits_{j=1}^{n}\Omega_q\sin(\mu_q t)\sigma_j^z + F(t)(a+a^\dagger), \rho \right] \hat{a}^\dagger\hat{a}\right] \nonumber\\
&=& -i\Tr\left[\left( \sum\limits_{j=1}^{n} g(a^\dagger \sigma_j + a\sigma_j^+)\rho -\rho \sum\limits_{j=1}^{n} g(a^\dagger \sigma_j + a\sigma_j^+) + F(t)(a+a^\dagger)\rho - \rho F(t)(a+a^\dagger) \right)\hat{a}^\dagger\hat{a} \right] \nonumber\\
&=& -i\Tr\left[\sum\limits_{j=1}^{n} g\left(\sigma_j\rho a^\dagger a a^\dagger - \sigma_j^+\rho a^\dagger aa
-\rho a^\dagger \sigma_j a^\dagger a - \rho a \sigma_j^+ a^\dagger a\right) 
+ F(t)\rho\left(\rho a^\dagger a + \rho a^\dagger a a^\dagger a - \rho a a^\dagger a - \rho a^\dagger a^\dagger a \right) \right] \nonumber\\
&=& -i\Tr\left[\sum\limits_{j=1}^{n} g\rho\left(\sigma^+\left(a^\dagger a - aa^\dagger \right)a + \sigma
a^\dagger \left(aa^\dagger - a^\dagger a \right) \right) +F(t)\rho\left(\left( a^\dagger a - a a^\dagger\right)a + a^\dagger\left(aa^\dagger - a^\dagger a \right)\right) \right]\nonumber\\
&=& -i\Tr\left[\sum\limits_{j=1}^{n} g\rho\left(-\sigma^+a + \sigma
a^\dagger \right) +\Omega_c\sin(\nu_c t)\rho\left(-a + a^\dagger\right) \right]\nonumber\\
&=& \left\langle \sum\limits_{j=1}^{n} ig\left(-\sigma^+a + \sigma
a^\dagger \right)\right\rangle +F(t)\left\langle i(-a + a^\dagger)\right\rangle \nonumber\\
&=& \sum\limits_{j=1}^{n}g\left\langle  i\left(-\sigma^+a + \sigma
a^\dagger \right)\right\rangle - F(t)\left\langle \hat{P}\right\rangle 
\end{eqnarray}

\begin{eqnarray}
\mathcal{S}_2 &=& \Tr\left[\Gamma_c\left(a\rho a^\dagger - \frac{1}{2}\left(a^\dagger a\rho + \rho a^\dagger a \right)\right)a^\dagger a  +\Gamma_q\left(\sigma \rho \sigma^+ - \frac{1}{2}\left(\sigma^+ a + \rho \sigma^+ \sigma \right)\right)a^\dagger a \right] \nonumber \\
&=& \Tr\left[\Gamma_c \rho\left(a^\dagger a^\dagger a a - \frac{1}{2}a^\dagger aa^\dagger a - \frac{1}{2}
a^\dagger aa^\dagger a\right) \right]\ \nonumber \\
&=& \Tr\left[\Gamma_c \rho\left(a^\dagger (aa^\dagger - 1) a - \frac{1}{2}a^\dagger aa^\dagger a - \frac{1}{2}
a^\dagger aa^\dagger a\right) \right]\ \nonumber \\
&=&-\Gamma_c \left\langle (a^\dagger a) \right\rangle
\end{eqnarray}

so the derivative of expectation photon number is 
\begin{equation}
\frac{d\langle \hat{a}^\dagger\hat{a} \rangle}{dt}= \sum\limits_{j=1}^{n}g\left\langle  i\left(-\sigma^+a + \sigma
a^\dagger \right)\right\rangle - F(t)\left\langle \hat{P}\right\rangle-\Gamma_c \left\langle (a^\dagger a) \right\rangle
\end{equation}
The input and output is 
\begin{equation}
\langle \hat{I}\rangle= -\langle \hat{P}\rangle
\end{equation}
\begin{equation}
\langle \hat{O}\rangle=\frac{d\langle \hat{a}^\dagger\hat{a} \rangle}{dt}+\Gamma_c \left\langle (a^\dagger a) \right\rangle
\end{equation}
and the relation between input and output
\begin{equation}
\hat{O} = F(t)\hat{I}
\end{equation}
where $F(t)$ is $\Omega_c\left((1-\sin(\cos\mu_c t)\right)$, due to the auxiliary qubits being one part of the environment, we discard the second term of the output, now the output is
\begin{equation}
\langle \hat{O}\rangle=\frac{d\langle \hat{a}^\dagger\hat{a} \rangle}{dt} +\Gamma_c \left\langle (a^\dagger a) \right\rangle + \mathcal{G}(t)
\end{equation}
with $\mathcal{G}(t)= \sum\limits_{j=1}^{n}g\left\langle  i\left(-\sigma^+a + \sigma a^\dagger \right)\right\rangle$. As $\mathcal{G}(t)$ depends on the interaction between the qubits and the cavity, it can be controlled by the external driving over the set of auxiliary qubits, which means that $G(t)$ will be close to zero then the qubits are off of resonance with the cavity.


\end{document}